\documentclass[fleqn,useAMS,usenatbib,usegraphicx]{mn2e}
\usepackage{times}
\usepackage{natbib}
\usepackage{epsfig}
\usepackage{rotating}
\usepackage{lscape}
\usepackage{amssymb}

\begin{document}

\def\kmm#1  {{\bf [KMM:~ #1]~}}
\def\new#1 {{\bf #1 }}
\def\cut#1 {\sout{#1} }

\newcommand{\htwo}{{\rm H}_2}
\newcommand{\pmn}{PMN\,J0134$-$0931}
\newcommand{\pks}{PKS\,1413+135}
\newcommand{\al}{\ensuremath{\alpha}}
\newcommand{\dal}{\ensuremath{\lsb \Delta \alpha/ \alpha \rsb}}
\newcommand{\gp}{\ensuremath{{g}_{p}}}
\newcommand{\lqcd}{\ensuremath{\Lambda_{\rm QCD}}}
\newcommand{\y}{\ensuremath{{m}_{e}/{m}_{p}}}
\newcommand{\dmu}{\ensuremath{\lsb \Delta \mu/\mu \rsb}}
\newcommand{\beq}{\begin{equation}}
\newcommand{\eeq}{\end{equation}}
\newcommand{\noi}{\noindent}
\newcommand{\lb}{\left(}
\newcommand{\rb}{\right)}
\newcommand{\lsb}{\left[}
\newcommand{\rsb}{\right]}
\newcommand{\hi}{H{\sc i}}
\newcommand{\kms}{km~s$^{-1}$}

\title[Constraining $\Delta \mu/\mu$ using CH$_3$OH lines]{Constraints on changes in the proton-electron mass ratio using methanol lines}

\author[Kanekar et al.]{N. Kanekar$^1$\thanks{E-mail: nkanekar@ncra.tifr.res.in; Swarnajayanti Fellow}
W. Ubachs$^2$,
K. M. Menten$^3$,
J. Bagdonaite$^3$,
A. Brunthaler$^3$,
C. Henkel$^{3,4}$,
\newauthor
S. Muller$^5$,
H. L. Bethlem$^2$,
M. Dapr{\`a}$^2$\\
$^1$National Centre for Radio Astrophysics, TIFR, Ganeshkhind, Pune - 411007, India\\
$^2$Department of Physics and Astronomy, VU University Amsterdam, De Boelelaan 1081, 
1081 HV Amsterdam, The Netherlands\\
$^3$Max-Planck-Institut f\"{u}r Radioastronomie, Auf dem H\"{u}gel 69, 53121, Bonn, Germany\\
$^4$Astronomy Department, King Abdulaziz University, PO Box 80203, 21589, Jeddah, Saudi Arabia\\
$^5$Department of Earth and Space Sciences, Chalmers University of 
Technology, Onsala Space Observatory, SE-43992 Onsala, Sweden
}

\maketitle

\begin{abstract}
We report Karl G. Jansky Very Large Array (VLA) absorption spectroscopy in four 
methanol (CH$_3$OH) lines in the $z = 0.88582$ gravitational lens towards PKS1830$-$211.
Three of the four lines have very different sensitivity coefficients $K_\mu$ to changes in 
the proton-electron mass ratio $\mu$; a comparison between the line redshifts 
thus allows us to test for temporal evolution in $\mu$. We obtain a stringent statistical 
constraint on changes in $\mu$ by comparing the redshifted 12.179~GHz and 60.531~GHz lines, 
$\dmu \leq 1.1 \times 10^{-7}$ ($2\sigma$) over $0 < z \leq 0.88582$, a factor of $\approx 2.5$ more 
sensitive than the best earlier results. However, the higher signal-to-noise ratio (by a factor 
of $\approx 2$) of the VLA spectrum in the 12.179~GHz transition also indicates that this 
line has a different shape from that of the other three CH$_3$OH lines (at $> 4\sigma$ 
significance). The sensitivity of the above result, and that of all earlier CH$_3$OH studies, 
is thus likely to be limited by unknown systematic errors, probably arising due to the 
frequency-dependent structure of PKS1830$-$211. A robust result is obtained by combining 
the three lines at similar frequencies, 48.372, 48.377 and 60.531~GHz, whose line profiles 
are found to be in good agreement. This yields the $2\sigma$ constraint $\dmu \lesssim 4 \times 
10^{-7}$, the most stringent current constraint on changes in $\mu$. We thus find no 
evidence for changes in the proton-electron mass ratio over a lookback 
time of $\approx 7.5$~Gyrs.

\end{abstract}


\section{Introduction} 
\label{sec:intro}

Astronomical spectroscopy in redshifted spectral lines has long been known to provide 
a probe of changes in the fundamental constants of physics (e.g. the fine 
structure constant $\alpha$, the proton-electron mass ratio $\mu \equiv m_p/m_e$, etc.), 
over large fractions of the age of the Universe \citep[e.g.][]{savedoff56}. Such 
temporal evolution is a generic prediction of field theories that attempt to unify the 
Standard Model of particle physics and general relativity \citep[e.g.][]{marciano84,damour94}.
The exciting possibility of low-energy tests of such unification theories has inspired 
a number of methods to probe fundamental constant evolution on a range of timescales, 
\citep[see, e.g.,][for a recent review]{uzan11}. Most of these methods, both in the laboratory 
and at cosmological distances, have been sensitive to changes in the fine structure 
constant $\alpha$ \citep[e.g.][]{webb01,peik04,gould06,rosenband08,molaro13}. 
However, fractional changes in $\mu$ are expected to be far larger than those in 
$\alpha$ in most theoretical scenarios, by factors of $10-500$ \citep[e.g.][]{calmet02,langacker02}. 

For many years, ultraviolet ro-vibrational molecular hydrogen ($\htwo$) lines provided 
the only technique to probe changes in $\mu$ on Gyr timescales 
\citep[][]{thompson75,varshalovich93,ubachs07}. The resulting sensitivity to $\dmu$ has 
been limited by the paucity of redshifted $\htwo$ absorbers \citep[e.g.][]{noterdaeme08}, 
systematic effects in the wavelength calibration of optical spectrographs 
\citep[][]{griest10,whitmore10,molaro13,rahmani13}, and the low sensitivity of $\htwo$ lines
to changes in $\mu$. The best limits on fractional changes in $\mu$ from this
technique are $\dmu \lesssim 10^{-5}$ ($2\sigma$) at redshifts $0 < z \lesssim 3$ 
\citep[e.g.][]{king11,vanweerdenburg11,rahmani13,bagdonaite14,vasquez14}.

The situation has changed dramatically in recent years with the development of new 
techniques using redshifted radio lines from different molecular species 
\citep[e.g.][]{darling03,chengalur03,flambaum07b,jansen11,levshakov11}. While the 
number of cosmologically-distant radio molecular absorbers is even smaller than the 
number of high-$z$ $\htwo$ absorbers \citep[just 5 radio systems; ][]{wiklind94,wiklind95,wiklind96,wiklind96b,kanekar05}, the high sensitivity of tunneling transition 
frequencies  in ammonia \citep[NH$_3$; ][]{vanveldhoven04,flambaum07b} and methanol 
\citep[CH$_3$OH; ][]{jansen11,levshakov11} to changes in $\mu$ has resulted in our 
best present constraints on changes in any fundamental constant on cosmological 
timescales. For example, 
\citet{kanekar11} obtained $\dmu < 2.4 \times 10^{-7}$ ($2\sigma$) from a comparison between the 
redshifts of CS, H$_2$CO and NH$_3$ lines from the $z \approx 0.685$ system towards 
B0218+357, while \citet{bagdonaite13b} 
obtained $\dmu \leq 2.6 \times 10^{-7}$ ($2\sigma$) using CH$_3$OH lines at 
$z \approx 0.88582$ towards PKS1830$-$211.

The best techniques to probe fundamental constant evolution are those that use spectral 
lines from a {\it single species}, of similar excitation and frequency
\citep[e.g. $\htwo$, OH, Fe{\sc ii}, CH$_3$OH;][]{thompson75,darling03,chengalur03,kanekar04a,kanekar04b,quast04,jansen11}. 
The different lines are then likely to arise in the same gas, implying that local velocity 
offsets should not be a source of systematic effects. The CH$_3$OH technique is especially 
interesting because CH$_3$OH has many strong radio lines with different frequency dependences 
on $\mu$; different line combinations thus provide independent probes of any evolution 
\citep{jansen11,levshakov11}. In this {\it Letter}, we report a robust new constraint 
on changes in $\mu$ from CH$_3$OH spectroscopy of the $z = 0.88582$ gravitational lens towards 
PKS1830$-$211 with the Karl G. Jansky Very Large Array (VLA).


\section{Observations, data analysis and spectra}
\label{sec:data}

The VLA observations of the methanol lines at $z = 0.88582$ towards PKS1830$-$211 were 
carried out in 2012 July and August, in the B-configuration (proposal 12A-389). Four 
CH$_3$OH lines were targeted, the 
$2_0-3_{-1}~E$, $1_0-0_0~A^+$ and $E$, and $1_0-2_{-1}~E$ transitions at $12.178597(4)$~GHz,
$48.3724558(7)$~GHz, $48.376892(10)$~GHz, and $60.531489(10)$~GHz,  respectively \citep[for the line
frequencies, see][and 
references therein]{muller04}. These are redshifted to observing frequencies of 
$\approx 6.46$~GHz (C-band), $25.65$~GHz (two lines; K-band) and $32.10$~GHz (Ka-band), respectively. 
The VLA C-band observations covering the redshifted 12.179~GHz line were carried 
out on July 13 and 14 (total on-source time: 7~hours), with both the K- and Ka-band observations 
on August 12, with on-source times of 1.5~hours (Ka-band) and 0.5~hours (K-band, covering 
both redshifted 48~GHz lines).

We note, in passing, that the 12.179~GHz line has the largest sensitivity coefficient 
($K_\mu \approx -32$) to changes in $\mu$, but is also the weakest of the CH$_3$OH lines that 
has been detected in the $z=0.88582$ absorber \citep[][]{bagdonaite13b}. All earlier 
studies using the CH$_3$OH lines to probe changes in $\mu$ have been limited by the low 
signal-to-noise ratio in this transition \citep[e.g.][]{bagdonaite13b,bagdonaite13}. 
We hence chose to obtain a deep spectrum in this line, spending the maximum integration time 
on its observations. 

Bandwidths of 8~MHz (C-band) and 32~MHz (K- and Ka-bands) were used in each of two 
intermediate frequency (IF) bands, which were further sub-divided into eight digital 
sub-bands, each with 128~channels. The two IF bands were offset from each other by half a 
digital sub-band, to ensure uniform sensitivity across the final spectra. 
Observations of 3C286, J1733$-$1304 (K- and Ka-bands) and J1229+0203 (C-band) were used to 
calibrate the flux density scale and the antenna bandpasses. No phase calibrator was 
observed, as PKS1830$-$211 is bright enough for self-calibration. Online Doppler tracking was not used.

All data were analysed in ``classic'' {\sc aips}, following standard procedures. 
For each band, after initial data editing, and flux and bandpass calibration, a number of 
absorption-free PKS1830$-$211 channels were averaged to form a ``channel-0'' dataset. A standard 
self-calibration procedure was then used to determine the antenna-based gains, until the 
gains and the image converged. The tasks {\sc uvsub} and {\sc uvlin} were used to 
subtract PKS1830$-$211's continuum from the calibrated visibilities, and
{\sc cvel} then used to shift the residual visibilities to the heliocentric frame. 
Each dataset was imaged to produce a spectral cube, with natural weighting used 
to maximise the sensitivity. This yielded cubes with angular resolutions of 
$\approx 0.4'' \times 0.2''$~(Ka-band), $0.66'' \times 0.34''$~(K-band) and 
$2.1'' \times 1.1''$~(C-band); the two components of PKS1830$-$211 \citep[separated 
by $\approx 1''$; ][]{rao88} were thus marginally resolved at K- and Ka-band. Finally,
the CH$_3$OH spectra were obtained via a cut through the location of the S-W source 
component against which the strong molecular absorption is seen 
\citep[e.g.][]{frye97,muller06}. At each frequency, the spectra from the two 
IF bands were combined to obtain the final spectrum. The C-band spectra from the two
runs were combined using weights based on their root-mean-square (RMS) noise values.

The final VLA CH$_3$OH optical depth spectra are shown in Fig.~\ref{fig:spectra}.
The optical depth RMS noise values against the S-W source component are $0.0062$ 
per $0.58$~\kms\ (Ka-band), $0.0048$ per $0.73$~\kms\ (K-band), and $0.00041$ per 
$0.73$~\kms\ (C-band). These optical depth estimates assume that 40\% of the flux 
density of PKS1830$-$211 is in the S-W component. For comparison, the Effelsberg spectra 
of \citet{bagdonaite13,bagdonaite13b} had optical depth RMS noise values of 
$0.0033$~per $1.8$~\kms\ (Ka-band), $0.0013$ per $2.3$~\kms\ (K-band) and 
$0.00040$ per 2.3~\kms\ (C-band). As noted earlier, the best current analyses 
\citep{bagdonaite13,bagdonaite13b} have been limited by the sensitivity in the 
redshifted 12.179~GHz transition. The sensitivity of the VLA C-band spectrum in 
this transition is a factor of $\approx 1.8$ better than that of the best earlier 
spectra.

\setcounter{figure}{0}
\begin{figure}
\centering
\includegraphics[scale=0.4]{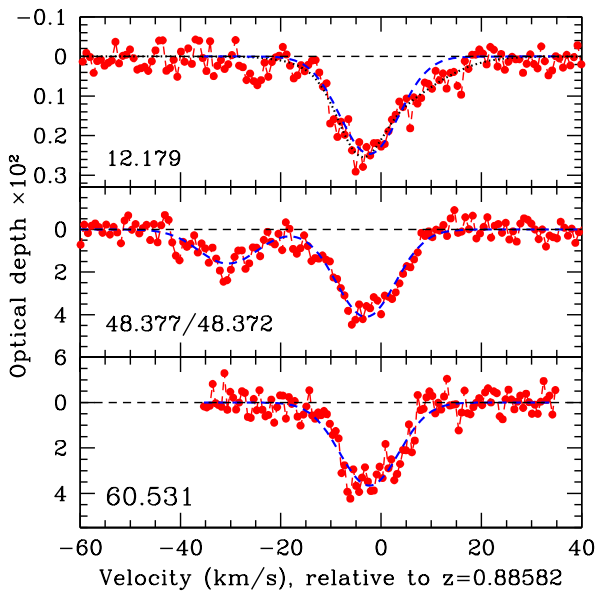}
\caption{VLA CH$_3$OH spectra at $z=0.88582$ towards PKS1830$-$211, with optical depth against the 
S-W image plotted 
against velocity, relative to $z=0.88582$ (heliocentric). The three panels show the 60.531~GHz 
line (bottom), the 48.372 and 48.377~GHz lines (middle), and the 12.179~GHz line (top); 
the velocity scale in the middle panel is for the 48.372~GHz line. The dashed curves in the lower 
two panels indicate the best-fit 1-Gaussian model that was obtained from a simultaneous fit to the 
60.531, 48.372 and 48.377~GHz lines, with the line FWHMs tied together. The dashed curve in the 
top panel shows the above (scaled) 1-Gaussian model overlaid on the spectrum in the 12.179~GHz 
transition; this model does not appear to provide a good fit to the weak absorption wing that 
extends beyond $\approx 10$~\kms. The dotted curve in the top panel shows a 2-Gaussian model 
that gives a good fit to the spectrum in the 12.179~GHz line.
\label{fig:spectra}}
\end{figure}

\section{Probing fundamental constant evolution}
\label{sec:alpha}

\setcounter{table}{0}
\begin{table*}
\begin{centering}
\label{table:fit} 
\begin{tabular}{|c|c|c|c|c|c|}
\hline
&&&&& \\ 
Transition        & Rest frequency  & $K_\mu$ & Redshift        & FWHM          & Peak optical depth \\
 	          &    GHz          &         &                 & \kms\ &  \\
\hline
&&&&& \\ 
$2_{-1} - 1_0$~E  & $60.531489(10)$ & $-7.4$  & $0.8858052(19)$ &               & $0.0433(15)$ \\
$0_0 - 1_0$~A$^+$ & $48.3724558(7)$ & $-1.0$  & $0.8858010(14)$ & $13.85(0.39)$ & $0.0566(18)$ \\
$0_0 - 1_0$~E     & $48.376892(10)$ & $-1.0$  & $0.8858010(14)$ &  	        & $0.01893(66)$ \\

&&&&& \\ 
\hline
\end{tabular}
\caption{Parameters of the best single-component Gaussian fit to the 60.531~GHz, 48.372~GHz,
and 48.377~GHz lines. The second and third columns give the line rest frequencies \citep[as measured 
in the laboratory; ][]{heuvel73,breckenridge95,muller04} and the sensitivity coefficents to changes
in $\mu$ \citep[][]{jansen11}. 
The redshifts of the 48.372 and 48.377~GHz lines were tied together in the fit (as these 
transitions have the same $K_\mu$ value), as were the column densities in the 60.531 and 
48.377~GHz lines (both E-type transitions). All three line FWHMs were also assumed to be 
the same. The measured redshifts, listed in column~(4), are in the heliocentric frame.}
\end{centering}
\end{table*}

A critical question in using multiple spectral lines to probe fundamental constant 
evolution is whether or not the lines arise in the same gas. This is important because 
any measured differences in the line redshifts might arise due to local velocity 
offsets in the absorbing gas. Unfortunately, it is not guaranteed that different 
transitions arise in the same gas, even for absorption lines from the same species, 
especially for gravitational 
lenses like the $z = 0.88582$ system towards PKS1830$-$211. For example, time variability in 
the structure of the background source could cause differences in the paths traced by different 
transitions, if the lines are not observed simultaneously \citep[e.g.][]{muller08}. Conversely, 
frequency-dependent 
structure in the background source could cause absorption at different frequencies (even 
if observed simultaneously) to probe different sightlines through the absorbing gas 
\citep[e.g.][]{martividal13,bagdonaite13b}. In the case of PKS1830$-$211, the S-W source 
component is known to be scatter-broadened, and hence more extended, at low frequencies 
\citep[e.g.][]{jones96}, due to which the C-band and K-/Ka-band observations may trace 
slightly different gas.

A basic test of whether different lines arise in the same gas is whether or not they have 
the same velocity structure. This can be simply tested by fitting the same template to 
the line profiles and checking whether the fit parameters (except for the amplitude)
are in agreement. A multi-Gaussian template profile was used for this purpose, with 
independent fits to the redshifted 12.179~GHz and 60.531~GHz lines, and a joint fit 
to the redshifted $48.372$~GHz and $48.377$~GHz lines (which are blended). The 
60.531~GHz, 48.372~GHz and 48.377~GHz lines were all found to be well fit by a
single Gaussian component (with reduced chi-square values $\chi^2_\nu \approx 1$), with the 
line full widths at half maximum (FWHMs) in good agreement (within $2\sigma$ significance). 
However, the single Gaussian fit to the 12.179~GHz line yielded a line FWHM that was larger 
than the FWHM of the other lines, at $4.3\sigma$ significance. Further, this fit yielded 
$\chi^2_\nu = 1.3$, with statistically significant residuals, suggesting that additional 
components are needed to model the line. This is also visually apparent from 
Fig.~\ref{fig:spectra}, where the 12.179~GHz line extends to $\approx 18$~\kms, unlike
the 60.531~GHz and 48.372~GHz lines. Thus, the 60.531~GHz, 48.372~GHz and 48.377~GHz 
lines appear likely to arise from the same sightline (as their line FWHMs are in agreement), 
while the sightline of the 12.179~GHz transition is likely to be different.

The most likely cause of this difference in sightlines is the frequency-dependent structure 
of PKS1830$-$211\footnote{The 1-month separation between the C-band
and K-/Ka-band observations may also contribute to the differences, especially given
that there was a gamma-ray flare in PKS1830$-$211 in May--July 2013 \citep{martividal13}.}, 
due to the larger scatter-broadening of the S-W source component at low frequencies 
\citep[e.g.][]{jones96}.
This would account for the larger velocity spread in the lower-frequency C-band transition, 
as this sightline would intersect a larger transverse region through the 
absorber, and hence, a more extended velocity field. Note that the size of the S-W radio 
emission changes dramatically with frequency, from $> 50$~mas at 2.3~GHz to $<10$~mas at 
8.4~GHz, with 6.46 GHz, the redshifted 12.179~GHz line frequency, in between 
\citep{guirado99}. The size diminishes even further at higher frequencies: at 14.5, 23.2 and 
43~GHz, almost all the S-W emission is concentrated in a compact source 
with FWHM~$< 0.5$~mas, representing the image of the AGN core without scatter broadening
\citep{jin03,sato13}. At 
$z = 0.88582$, 1~mas corresponds to 8~pc, implying that we are probing scales of $\sim 100$~pc 
(i.e. the size of a typical Giant Molecular Cloud, GMC) at $6.46$~GHz, but only $\sim 5$~pc 
(i.e. a small part of a GMC) at $32.10$~GHz. This makes velocity differences of a few km~s$^{-1}$ 
and different line shapes certainly plausible. The above issue was considered by 
\citet{bagdonaite13b}, who assumed that it could cause a shift of $\approx 1$~\kms\ between 
the different lines \citep{martividal13}; unfortunately, it is difficult to directly estimate 
the systematic error here. It thus appears that the 12.179~GHz line should not be compared
with the higher-frequency CH$_3$OH lines to probe changes in $\mu$. The other three lines, 
closer in frequency, are likely to arise from the same gas and hence appear to be well-suited for 
use in probing fundamental constant evolution. However, since the difference in line FWHMs has only 
$\approx 4.3\sigma$ significance, we provide below results based on both analyses, i.e. both 
including and excluding the 12.179~GHz line.

We used two independent approaches to test for changes in the proton-electron mass ratio:
(1)~a joint multi-Gaussian fit to the 60.531~GHz, 48.372~GHz and 48.377~GHz lines, and 
(2)~cross-correlation of individual pairs of lines. The advantage of the former is that it
implicitly tests whether all the lines can be fit with the same template (i.e. the same 
number of Gaussian components, with the same FWHMs). Furthermore, the joint fit 
simultaneously makes use of all the information in the different lines 
\citep[e.g.][]{webb01,molaro13,rahmani13}. 
Conversely, the advantage of the cross-correlation approach is that it is non-parametric, 
making no assumptions about the line shapes or the number of components that must be fit 
to the lines (which can affect the results, especially for complex line profiles)
\citep[e.g.][]{kanekar04b,kanekar10b,levshakov12}.

In the first approach, the package {\sc vpfit} was used to carry out a simultaneous multi-Gaussian 
fit to the 60.531~GHz, 48.372~GHz and 48.377~GHz lines, aiming to minimize $\chi^2_\nu$ by varying 
the fit parameters\footnote{Note that little is known about the CH$_3$OH hyperfine structure which 
has been ignored here. For a few lines, \citet{heuvel73} find the hyperfine structure extends over 
$\lesssim 50$~kHz, i.e. smaller than our channel spacing; it should have a negligible effect
on our results.}
. The velocity structure was assumed to be the same in all the lines, with the 
same number of components and the same velocity widths (which were hence assumed to be the same 
in the fitting process). This essentially assumes that all the lines arise in the same absorbing 
gas. Since the 48.372~GHz and 48.377~GHz lines have the same sensitivity coefficient to changes 
in $\mu$ ($K_\mu = -1$), these line redshifts were also assumed to be the same. Finally, the 
60.531~GHz and 48.377~GHz lines are both E-type lines (the 48.372~GHz line is an A-type); local 
thermal equilibrium was assumed for the ratios of the strengths of the E-type lines.

A single component Gaussian model was found to yield an excellent fit to all three lines, with 
$\chi^2_\nu = 1.07$, and noise-like residuals (via a Kolmogorov-Smirnov rank-1 test) in each 
spectrum, after subtracting out the fitted profiles. Five free parameters (listed in Table~1) were 
used in the fit, two column densities for the E-type and A-type lines (see above), a single FWHM for 
all lines, and two redshifts for the 60.531~GHz line and the pair of 48~GHz lines. 
The velocity offset between the redshifts of the 60.531~GHz line ($z_A$) 
and the 48~GHz pair ($z_B$) is given by $\Delta V/c$ = $(z_A - z_B)/(1 + {\bar z})$, where ${\bar z}$ 
is the average of the two redshifts. This yields $\Delta V= (+0.66 \pm 0.39)$~\kms, i.e. 
$\dmu = (\Delta V/c) \times (K_{\mu,A} - K_{\mu,B}) = (-3.5 \pm 2.0) \times 10^{-7}$, 
where $K_{\mu,A}=-7.4$ and $K_{\mu,B} = -1$ are the sensitivity coefficients of the 60.531~GHz
line and the 48~GHz pair to changes in $\mu$. Similar results were obtained on removing 
the assumption that the two 48~GHz lines (A- and E-types) arise at the same redshift. 

The second approach was carried out on individual pairs of lines using two non-parametric schemes, 
the cross-correlation method \citep{kanekar04b} and the ``sliding-distance'' method 
\citep{levshakov12}. These yielded very similar results, so we will only discuss the cross-correlation 
approach in detail. The cross-correlation of the two strongest lines, at 60.531~GHz and 
48.372~GHz, yielded a velocity offset of $\Delta V = 
(+0.46 \pm 0.43)$~\kms, with the 60.531~GHz line at higher velocities. The RMS error was estimated by 
cross-correlating $10^4$ pairs of simulated spectra, obtained by adding independent representations 
of Gaussian noise to the best-fit profiles, with the noise spectra characterized by the RMS noise 
values of the observed spectra. This yields $\dmu = (-2.4 \pm 2.3) \times 10^{-7}$, similar 
to the result obtained earlier from the Gaussian-fitting analysis. The slightly higher error in 
the cross-correlation approach is because only the two strongest lines were used here, so all the 
information available in the joint Gaussian-fitting has not been used. Further, the entire velocity 
range was not used in the cross-correlation, with the velocity interval containing a blend of 
the 48.372~GHz and 48.377~GHz lines excluded from the analysis, to reduce systematic errors.

The analysis using the 12.179~GHz line was restricted to the cross-correlation approach. 
This also takes into account the possibility that the 12.179~GHz line might not be offset in 
velocity from the other lines, but might merely have additional spectral components, in addition 
to the strongest component seen in the other lines. A cross-correlation analysis was hence carried 
out on the 12.179~GHz and 60.531~GHz pair, following the approach detailed above. This yielded 
a velocity offset of $\Delta V = (+0.22 \pm 0.43)$~\kms, again with the 60.531~GHz line at higher 
velocities, giving $\dmu = (-2.9 \pm 5.7) \times 10^{-8}$. Again, no evidence is seen for 
changes in $\mu$ with cosmological time.


\section{Discussion and Conclusions}
\label{sec:discussion}

The $z = 0.88582$ absorber towards PKS1830$-$211 is the only object so far with a 
detection of redshifted CH$_3$OH absorption \citep[][]{muller11} and has hence been the focus of 
tests of changes in the proton-electron mass ratio using these lines. Initially, 
\citet{ellingsen12} obtained $\dmu \leq 4.2 \times 10^{-7}$ (at $2\sigma$ significance), 
via a comparison between the redshifts of the 12.179~GHz and the 60.531~GHz lines. Later, 
\citet{bagdonaite13} improved this to $\dmu \leq 2 \times 10^{-7}$, using the 12.179~GHz, 
48.372~GHz, 48.377~GHz and 60.531~GHz lines. Most recently, \citet{bagdonaite13b} 
combined the 12.179~GHz line with nine other lines, at rest frequencies $\leq 492.279$~GHz, 
and also carried out a multi-dimensional regression analysis to include systematic effects 
in their error budget, to obtain $\dmu \leq 2.6 \times 10^{-7}$ $(2\sigma)$ 
(combining statistical and systematic errors in quadrature). In the present work,
we have combined the 12.179~GHz line with the 60.531~GHz line to obtain a constraint 
of high apparent sensitivity, $\dmu \leq 1.1 \times 10^{-7}$ ($2\sigma$), a factor of 
$\approx 2.5$ better than the best earlier results \citep[][]{bagdonaite13b}. 

All the earlier analyses used independent fits to the different lines and did not 
directly test whether the lines arise in the same gas. The 12.179~GHz line 
was a critical component of all analyses, especially because this line has $K_\mu = -32$, 
one of the largest sensitivity coefficients of the CH$_3$OH lines. Including this line in 
the comparison thus gives a large lever arm (i.e. a large $\Delta K_\mu$) in tests of 
changes in $\mu$. Unfortunately, this is one of the weaker CH$_3$OH lines, and its spectra 
have had the lowest signal-to-noise ratios of the lines that have been used so far. As 
noted earlier \citep{bagdonaite13b}, this is also the line that is most likely to be affected
by differing sightlines, due to the large scatter broadening of the S-W source component 
of PKS1830$-$211 at the low line frequency. The relatively low sensitivity of the earlier 
spectra in this transition has meant that it has not been possible to test whether the line 
shape is the same as that of the other CH$_3$OH lines.

Our new, higher-sensitivity, VLA spectrum indicates that the 12.179~GHz line profile is indeed
different from the other three line profiles: the line FWHM is larger, at $\approx 4.3\sigma$ 
significance, and the 1-Gaussian model yields a relatively high reduced chi-square value, 
$\chi_\nu^2 \approx 1.3$. This suggests that the sightline in the 12.179~GHz 
transition traces different absorbing gas from that detected in the other three lines. If so, 
this implies that analyses that include the 12.179~GHz transition (i.e. all earlier results 
from the $z = 0.88582$ absorber, as well as our result above) would incur unknown systematic 
errors due to local velocity structure in the gas. 

We hence recommend that the 12.179~GHz line should not be used with the high-frequency 
CH$_3$OH lines from the $z = 0.88582$ absorber to probe changes in the proton-electron 
mass ratio. The best probe of changes in $\mu$ using the CH$_3$OH lines is likely to arise 
from a combination of the 48.372/48.377~GHz pair and the 60.531~GHz line, since these lines 
both have a large difference between sensitivity coefficients ($\Delta K_\mu = 6.4$),
and lie at nearby observing frequencies, where the background source has a similar 
structure \citep[S-W component size~$< 0.5$~mas;][]{jin03,sato13}. The lines should be 
observed simultaneously, so that time variability in PKS1830$-$211 does not cause systematic 
errors. Finally, our current result from these lines, $\dmu \lesssim 4 \times 10^{-7}$ $(2\sigma)$,
is based on a fairly short VLA integration time ($2$~hrs). Deeper VLA spectroscopy in 
these lines will allow a significant improvement in the sensitivity to temporal changes in $\mu$.

In conclusion, our VLA observations of the CH$_3$OH lines from the $z = 0.88582$ 
absorber towards PKS1830$-$211 suggest that the shape of the 12.179~GHz line is different 
from that of the 48.372, 48.377 and 60.531~GHz lines. The FWHM of the 12.179~GHz line is 
different from that of the other lines at $\approx 4.3\sigma$ significance, and the best fit 
with a 1-Gaussian model has a relatively high reduced chi-square value, $\chi_\nu^2 \approx 1.3$, 
in contrast to similar fits to the other three lines, which yield $\chi_\nu^2 \approx 1$.
Including the 12.179~GHz line in the analysis to probe changes in $\mu$ gives a high 
statistical sensitivity to changes in $\mu$: $\dmu = (-2.9 \pm 5.7) \times 10^{-8}$, 
i.e. $\dmu \leq 1.1 \times 10^{-7}$ at $2\sigma$ significance. While this constraint 
on changes in $\mu$ is a factor of $\approx 2.5$ better than the best earlier results 
using CH$_3$OH lines, both it and all earlier CH$_3$OH analyses in the literature appear 
likely to be subject to unknown systematic errors due to the difference in the shapes 
of the 12.179~GHz line and the higher-frequency CH$_3$OH lines. The CH$_3$OH 48.372, 
48.377 and 60.531~GHz lines have shapes consistent with each other and are thus likely 
to arise in the same gas. Our most robust result, obtained via both joint Gaussian-fitting 
and cross-correlation approaches, stems from combining the three high-frequency 
lines: $\dmu \lesssim 4 \times 10^{-7}$ $(2\sigma)$. While this gives an apparently less 
stringent constraint than that 
from analyses that include the 12.179~GHz line, the fact that the lines used in the 
analysis have the same shape suggests that it is the most reliable of the present 
constraints on changes in $\mu$ based on CH$_3$OH spectroscopy. We thus find no evidence 
for changes in the proton-electron mass ratio 
out to $z = 0.88582$, with $\dmu \lesssim 4 \times 10^{-7}$ $(2\sigma)$ over a lookback 
time of $\approx 7.5$~Gyr.

\section*{Acknowledgments}

The National Radio Astronomy Observatory is a facility of the National Science Foundation, operated under 
cooperative agreement by Associated Universities, Inc. NK acknowledges support from the DST, India, via a Swarnajayanti Fellowship. 

\footnotesize{
\bibliographystyle{mn2e}
\bibliography{ms}
}

\label{lastpage}
\end{document}